\documentclass[aoas,MSNbibl,nameyear,dvips]{arximspdf}
\usepackage{multirow,dcolumn}
\usepackage{subeqn}
\usepackage{breakurl}

\doi{10.1214/13-AOAS664} 
\volume{7}
\issue{4}
\pubyear{2013}
\firstpage{1983}
\lastpage{2006}

\makeatletter
\newproclaim{example}{Example}
\newcolumntype{d}[1]{D{.}{.}{#1}}
\newcommand{\rrvert}{\vert}
\newcommand{\llvert}{\vert}
\makeatother

\begin{document}
\begin{frontmatter}

\title{Calibrated imputation of numerical data under linear edit
restrictions}
\runtitle{Calibrated imputation under edit restrictions}

\begin{aug}
\author[A]{\fnms{Jeroen} \snm{Pannekoek}\thanksref{m1}\ead[label=e1]{j.pannekoek@cbs.nl}},
\author[B]{\fnms{Natalie} \snm{Shlomo}\corref{}\thanksref{m2}\ead[label=e2]{natalie.shlomo@manchester.ac.uk}}
\and
\author[A]{\fnms{Ton} \snm{De Waal}\thanksref{m1}\ead[label=e3]{t.dewaal@cbs.nl}}
\runauthor{J. Pannekoek, N. Shlomo and T. De Waal}
\affiliation{Statistics Netherlands\thanksmark{m1} and  University of Manchester\thanksmark{m2}}
\address[A]{J. Pannekoek\\
T. De Waal\\
Statistics Netherlands\\
PO Box 24500\\
2490 HA The Hague\\
The Netherlands\\
\printead{e1}\\
\phantom{E-mail:\ }\printead*{e3}}

\address[B]{N. Shlomo\\
School of Social Sciences\\
University of Manchester\\
Humanities Bridgeford Street\\
Manchester, M13 9PL\\
United Kingdom\\
\printead{e2}}
\end{aug}

\received{\smonth{3} \syear{2012}}
\revised{\smonth{6} \syear{2013}}

\begin{abstract}
A common problem faced by statistical institutes is that
data may be missing from collected data sets. The typical way to overcome
this problem is to impute the missing data. The problem of imputing missing
data is complicated by the fact that statistical data often have to satisfy
certain edit rules and that values of variables across units sometimes have
to sum up to known totals. For numerical data, edit rules are most often
formulated as linear restrictions on the variables. For example, for data
on enterprises edit rules could be that the profit and costs of an
enterprise should sum up to its turnover and that the turnover should be at
least zero. The totals of some variables across units may already be known
from administrative data (e.g., turnover from a tax register) or estimated
from other sources. Standard imputation methods for numerical data as
described in the literature generally do not take such edit rules and
totals into account. In this article we describe algorithms for imputing
missing numerical data that take edit restrictions into account and ensure
that sums are calibrated to known totals. These algorithms are based on a
sequential regression approach that uses regression predictions to impute
the variables one by one. To assess the performance of the imputation
methods, a simulation study is carried out as well as an evaluation study
based on a real data set.
\end{abstract}


\end{frontmatter}

\section{Introduction}\label{sec1}

National statistical institutes (NSIs) publish figures on many aspects of
society. To this end, NSIs collect data on persons, households,
enterprises, public bodies, etc. A major problem arising from the data
collection is that data may be missing. Some units that are selected for
data collection cannot be contacted or may refuse to respond altogether.
This is called unit nonresponse. For many records, that is, the data of
individual respondents, data on some of the items may be missing. Persons
may, for instance, refuse to provide information on their income or on
their sexual habits, while at the same time giving answers to other, less
sensitive questions on the questionnaire. Enterprises may not provide
answers to certain questions, because they may consider it too complicated
or too time-consuming to answer these questions. Missing items of otherwise
responding units is called item nonresponse. Whenever we refer to missing
data in this article we will mean item nonresponse, unless indicated
otherwise.

Missing data is a well-known problem that has to be faced by basically all
institutes that collect data on persons or enterprises. In the statistical
literature ample attention is hence paid to missing data. The most common
solution to handle missing data in data sets is imputation, where missing
values are estimated and filled in. An important problem of imputation is
to preserve the statistical distribution of the data set. This is a
complicated problem, especially for high-dimensional data. For more on this
aspect of imputation and on imputation in general we refer to \citet{KalKas86}, \citet{Rub87}, \citet{KovWhi95}, \citet{Sch97},
\citet{LitRub02}, \citet{Lon05}, De Waal, Pannekoek and Scholtus
(\citeyear{DeWPanSch11}) and the references therein.

At NSIs the imputation problem is further complicated owing to the
existence of constraints in the form of edit restrictions, or edits for
short, that have to be satisfied by the data. Examples of such edits are
that the profit and the costs of an enterprise have to sum up to its
turnover, and that the turnover of an enterprise should be at least zero.
Records that do not satisfy these edits are considered incorrect.

Although some research on general approaches to imputation of numerical
data under edit restrictions has been carried out [see, e.g., Raghunathan,
Solenberger and Van Hoewyk (\citeyear{RagSolVan02}), \citet{Tem07}, \citet{Holetal10},
Coutinho, De Waal and Remmerswaal (\citeyear{CoudeWRem11}) and Chapter~9 in De Waal,
Pannekoek and Scholtus (\citeyear{DeWPanSch11})], this is a rather neglected area. The most
commonly used approach for numerical data under edit restrictions is
imputation based on a truncated multivariate normal model [see, e.g.,
\citet{Gew91}, \citet{Tem07} and De Waal, Pannekoek and Scholtus
(\citeyear{DeWPanSch11})]. An obvious drawback of basing imputations on a posited truncated
multivariate normal model is that this can only lead to good imputations
when the data approximately follow such a distribution. \citet{DraWin} have developed a balancing approach that allows several component
variables within the same record to add up to a total variable. Drawbacks
of that approach are that variables may be involved in at most one
balancing edit (see Section~\ref{sec2.2} for a definition of balancing edits), and
that in their implementation Draper and Winkler only use so-called ratio
imputation with a single predictor. More advanced imputation methods are
not considered. For categorical data under edit restrictions some work has
been done by Winkler (\citeyear{Win08N1,Win08N2}) and by Coutinho, De Waal and Shlomo
(\citeyear{CoudeWShl13}).

A further complication is that numerical data sometimes have to sum up to
known totals. As far as we are aware, the problem of imputing numerical
data subject to edit restrictions within records and population totals
across records has not yet been studied in the literature.

The purpose of the present article is to introduce techniques that can be
used to extend existing imputation methods for numerical data such that the
imputed data will satisfy edits and preserve population totals. The
imputation methods studied are based on a sequential regression approach
which means that the variables with missing values are imputed one after
another by using a regression model with (all) other variables as
predictors. Algorithms for (multiple) sequential regression imputation are
known as SRMI (Sequential Regression Multiple Imputation) and MICE
(Multiple Imputation by Chained equations) and are described by, for
example, \citet{Ragetal01}, Van Buuren and Groothuis-Oudshoorn
(\citeyear{VanGro11}) and \citet{Van12}. Sequential regression imputation can be
applied in different ways, depending on what the imputed data are to be
used for. The simplest way, often applied at NSIs, is to use the predicted
value directly as the imputation (predicted mean imputation), which is
suitable if interest is in (subgroup) means and totals. To better preserve
the variability in the data, this method can be extended by adding random
residuals to the predicted means.

The focus of this article is on modifications of the different sequential
regression techniques, from different statistical frameworks, that make
them applicable in our setting, that is, by satisfying edits and preserving
population constraints. Depending on one's goals and statistical framework,
a method can then be chosen that is best suited for the application at
hand.

An important issue is variance estimation after imputation. In this article
we will not go into the details of variance estimation, however, except for
a discussion in Section~\ref{sec7}.

The problem of imputing missing data while satisfying edits and preserving
totals can arise in the context of a survey among a subpopulation of
enterprises. Often large enterprises, for example, enterprises with a
number of employees exceeding a certain threshold value, are automatically
included in a survey. As already noted, some of those enterprises may,
however, not provide answers to all questions, and some may not answer any
question at all. Totals of some variables on the survey corresponding to
this subpopulation of enterprises may be known from other sources, for
example, from available register data, or may already have been estimated
from other sources. NSIs generally aim to publish a single figure for the
same phenomenon. One of the ways to achieve this is to benchmark data to
totals that are known or estimated from other sources. As data of
enterprises usually have to satisfy edits, imputation of such a data set
then naturally leads to the problem we consider in the present article.

In the case of a (nonintegral) sample survey with item nonresponse,
benchmarking to totals can either be done by first imputing the missing
data and then adjusting the sampling weights or by retaining the original
sampling weights and imputing so that totals are preserved. Our imputation
algorithms are a first step toward the latter approach. In Section~\ref{sec2} we
will elaborate more on this.

\citet{Rub76} introduced a classification of missing data mechanisms. He
distinguishes between Missing Completely At Random (MCAR), Missing At
Random (MAR) and Not Missing At Random (NMAR). Roughly speaking, in the
case of MCAR there is no relation between the missing data pattern, that
is, which data are missing, and the values of the data, either observed or
missing. In the case of MAR there is a relation between the missing data
pattern and the values of the observed data, but not between the missing
data pattern and the values of the missing data. Using the values of the
observed data, one can then correct for the relation between the missing
data pattern and the values of the observed data since within classes of
the observed data the missing data mechanism is MCAR again. In the case of
NMAR there is a relation between the missing data pattern and the values of
the missing data. Such a relation cannot be corrected for without positing
a model. Given that the missing data mechanism is either MCAR or MAR, we
can test whether the data are MCAR or MAR. However, there are no
statistical tests to differentiate between MCAR/MAR and NMAR. In practice,
the only way to distinguish MCAR/MAR from NMAR is by logical reasoning. For
more on missing data mechanisms we refer to \citet{LitRub02},
\citet{McKetal07} and \citet{Sch97}.

In this article we assume that the missing data mechanism is MCAR. Our
imputation methods can, however, easily be extended to the case of MAR by
constructing imputation classes within which the missing data mechanism is
MCAR.

Throughout this article we also assume that the missing data can indeed be
imputed in a manner consistent with the edits and the totals. This means we
assume that the data set to be imputed does not contain any remaining
errors. Such errors may have been found by automatic editing [see, e.g.,
\citet{FelHol76}] or other editing techniques [see De Waal,
Pannekoek and Scholtus (\citeyear{DeWPanSch11}) for an overview].

The remainder of this article is organized as follows. Section~\ref{sec2} introduces
the edit restrictions and sum constraints we consider in this article, and
explains the problem we consider in more detail. Section~\ref{sec3} develops two
sequential imputation algorithms for our problem. Section~\ref{sec4} develops a
third imputation algorithm. This algorithm is an extension of MCMC
algorithms used in multiple sequential regression imputation. It uses a
fully imputed data set satisfying edits and totals as a starting point and
aims to improve the statistical quality of the imputations. The fully
imputed data set satisfying edits and totals can, for example, be obtained
by one of the two sequential approaches developed in Section~\ref{sec3}. A
simulation study is described in Section~\ref{sec5} and an application on a real
data set in Section~\ref{sec6}. Finally, Section~\ref{sec7} concludes with a brief
discussion.

\section{Constraints on the imputed data}\label{sec2}

The problem addressed in this article can be described concisely as the
imputation of missing values in an $r \times n$ data matrix with $r$ the
number of rows (units) and $n$ the number of columns (variables), when the
imputed data in each row has to satisfy certain linear restrictions (row
restrictions) and the sums of some of the columns must equal known values
(column or sum constraints). In this section we describe in some detail how
this problem arises in the context of surveys or censuses with missing
data, edit rules and known population totals.

\subsection{Known population totals (column constraints)}\label{sec2.1}

In the usual sample survey setting, units are selected from a population
according to a specified sampling design. For an equal probability sample
of fixed size $s$ from a population of size $N$, all inclusion
probabilities are $s/N$. Estimates of (sub)population totals and other
parameters of interest are then calculated by using the sampling weights
that are the inverse of the inclusion probabilities. In particular, for the
total of a variable~$x_{j}$, this weighting estimator is $\hat{X}_{j} =
\sum_{i}^{s} \frac{N}{s}x_{ij}$, with $x_{ij}$ the value of $x_{j}$ for
unit $i$.

In practice, due to unit nonresponse, data are often only obtained for a
subset of the intended sample units and the (effective) sample size, or the
number of responding units, is $r < s$. A simple correction for unit
nonresponse is to use the effective sample size $r$ instead of the intended
sample size $s$ in this estimator, that is, by inflating the weights by the
inverse of the nonresponse fraction, $s / r$. If for some variables the
population totals are known, the weights can also be adjusted such that the
estimated totals for these variables equal their known values. Such weights
are said to be ``calibrated'' on the variables with known totals and are
not equal for all units [see, e.g., S\"{a}rndal and Lundstr\"{o}m (\citeyear{SarLun05})].
The effect of calibration on the weights is such that if an estimated total
is too low, the weights for units with low values for that variable will
decrease, whereas the weights for units with high values will increase.
Note that changing the weights will affect the estimates for all variables,
but this can be motivated by the observation that apparently the random
selection of the sample or unit nonresponse resulted in too many units with
a low value on this particular variable and adjusting the unit weights
corrects for this unbalanced selection of units. For large samples and
small unit nonresponse fractions calibration should have only minor effects
on the weights.

The situation with item nonresponse is different from unit nonresponse
because in this case the nonresponse fractions will vary greatly between
variables and, consequently, a simple nonresponse adjustment to the unit
level weights is not an option. The usual approach to deal with item
nonresponse is therefore to impute the missing values so that for the $r$
units a complete data set is obtained. Estimation weights, $N/r$ in the
equal probability case, will then be used that only reflect unit
nonresponse. When population totals are known, calibration of these weights
could again be used to ensure that estimates of totals will be equal to the
known values. However, for variables with imputed values, differences
between estimated totals and their known values are now caused not only by
an unfavorable realization of the random sample selection or selective unit
nonresponse, but also by systematic errors in the imputed values
(imputation bias). For large sample sizes and small unit nonresponse
fractions the difference between estimated and known population totals will
be mainly due to imputation bias. In such cases, it is not desirable to
adjust the weights by calibration because the adjustments do not correct
for an unbalanced selection of units but for imputation bias in specific
variables and there is no compelling reason to let this adjustment affect
the estimates of all other variables.

In this article we therefore consider to solve the inconsistency problem by
adjustment of the imputations that contribute to the inconsistent
estimates, but leave the weights unchanged so that the adjustments have no
effect on other variables.

For equal weights, the sum constraints on the estimates can be expressed as
$\hat{X}_{j} = \sum_{i = 1}^{r} \frac{N}{r}x_{ij} = X_{j}^{\mathrm{pop}}$, with
$X_{j}^{\mathrm{pop}}$ the known population total. In terms of the unweighted sample
totals, these constraints imply $\sum_{i = 1}^{r} x_{ij} =
\frac{r}{N}X_{j}^{\mathrm{pop}} = X_{j}$, say. Although in the application in this
article only equal weights are considered, in general weights will often be
unequal and the column constraints would be weighted sum constraints of the
form $\hat{X}_{j} = \sum_{i = 1}^{r} w_{i}x_{ij} = X_{j}^{\mathrm{pop}}$, with
$w_{i}$ the unit weights.

\subsection{Linear edit restrictions (row restrictions)}\label{sec2.2}

The edit restrictions imply within record (or row) restrictions on the
values of the variables. In this article we focus on linear edits for
numerical data. Linear edits are either linear equations or linear
inequalities. We assume that edit $k\ (k=1,\ldots,K)$ can be written in either
of the two following forms:
\renewcommand{\theequation}{\arabic{equation}a}
\begin{equation}
a_{1k}x_{1} + \cdots + a_{nk}x_{n} +
b_{k} = 0\label{eq1a}
\end{equation}
or
\renewcommand{\theequation}{\arabic{equation}b}
\setcounter{equation}{0}
\begin{equation}
a_{1k}x_{1} + \cdots + a_{nk}x_{n} +
b_{k} \ge 0. \label{eq1b}
\end{equation}

Here the $a_{jk}$ and the $b_{k}$ are certain constants, which define the
edit.

Edits of type (\ref{eq1a}) are referred to as balance edits. An example of such an
edit is
\renewcommand{\theequation}{\arabic{equation}}
\setcounter{equation}{1}
\begin{equation}
T = P + C,\label{eq2}
\end{equation}
where $T$ is the turnover of an enterprise, $P$ its profit and $C$ its
costs. Edit (\ref{eq2}) expresses that the profit and the costs of an enterprise
should sum up to its turnover. A record not satisfying this edit is
obviously incorrect. Edit (\ref{eq2}) can be written in the form (\ref{eq1a}) as $T - P - C
= 0$.

Edits of type (\ref{eq1b}) are referred to as inequality edits. An example is
%
%
\begin{equation}
T \ge 0,\label{eq3}
\end{equation}
expressing that the turnover of an enterprise should be nonnegative. One
has to take care that the edits are defined correctly as otherwise bias
might be introduced by making the data conform to incorrect edit rules.

\section{Sequential imputation algorithms satisfying edits and totals}\label{sec3}

In this section we present two algorithms for imputing data that satisfy
edits and totals. Both algorithms are sequential approaches based on
standard regression imputation techniques, but with (slight) adjustments to
the imputed values such that they satisfy edits and totals. Below we first
explain how a sequential approach can be used.

\subsection{Using a sequential approach}\label{sec3.1}

In order to be able to use a sequential approach, we apply Fourier--Motzkin
elimination ([\citet{Duf74}, De Waal, Pannekoek and Scholtus (\citeyear{DeWPanSch11})].
Fourier--Motzkin elimination is a technique to project a set of linear
constraints involving $q$ variables onto a set of linear constraints
involving $q-1$ variables. It is guaranteed to terminate after a finite
number of steps. The essence of Fourier--Motzkin elimination is that every pair of two
constraints, say, $L(x_{1}, \ldots,x_{r - 1},x_{r + 1}, \ldots,x_{q}) \le
x_{r}$ and $x_{r} \le U(x_{1}, \ldots,x_{r - 1},x_{r + 1}, \ldots,x_{q})$,
where $x_{r}$ is the variable to be eliminated and $L(x_{1}, \ldots,x_{r -
1},x_{r + 1}, \ldots,x_{q})$ and $U(x_{1}, \ldots,x_{r - 1},x_{r + 1},
\ldots,x_{q})$ are linear expressions in the other variables, leads to a
constraint $L(x_{1}, \ldots,x_{r - 1},x_{r + 1}, \ldots,\break  x_{q}) \le U(x_{1},
\ldots,x_{r - 1},x_{r + 1}, \ldots,x_{q})$ involving these other variables.
The main property of Fourier--Motzkin elimination is that the original set
of constraints involving $q$ variables can be satisfied if and only if the
corresponding projected set of constraints involving $q-1$ variables can be
satisfied. By repeated application of Fourier--Motzkin elimination, we can
derive an admissible interval for one of the values to be imputed. The main
property of Fourier--Motzkin guarantees that if we impute a value within
this admissible interval, the remaining values can be imputed in a manner
consistent with the constraints, that is, such that all constraints are
satisfied. Fourier--Motzkin elimination is closely related to the
Fellegi--Holt method [see \citet{FelHol76}] for automatically
detecting errors in a data set. A major difference is that in their article
Fellegi and Holt focus on categorical data instead of numerical data.
Moreover, in our article Fourier--Motzkin is only used to impute the data in
a manner consistent with the edits, not to find any errors in the data.

We now illustrate how we apply Fourier--Motzkin elimination. Say we want to
impute a variable $x_{j}$. We consider the records in which the value of
variable $x_{j}$ is missing. In order to impute a missing field
$x_{\mathit{ij}}$ in record~$i$, we first fill in the observed and
previously imputed values (if any) for the other variables in record $i$
into the edits. This leads to a reduced set of edits involving only the
remaining variables to be imputed in record $i$.

Next, we eliminate all equations from this reduced set of edits. That is,
we sequentially select any equation and one of the variables $x\ (x \ne
x_{j})$ involved in the selected equation. We then express $x$ in terms of
the other variables in the selected equation and substitute this expression
for $x$ into the other edits in which $x$ is involved. In this way we
obtain a set of edits involving only inequality restrictions for the
remaining variables. Once we have obtained imputation values for variables
involved in the set of inequalities, we find values for the variables that
were used to eliminate the equations by means of back-substitution.

From the set of inequality restrictions we eliminate any remaining
variables except $x_{ij}$ itself by means of Fourier--Motzkin elimination.\
Using Fourier--Motzkin elimination guarantees that the eliminated variables
can later be imputed themselves such that all edits become satisfied.

After Fourier--Motzkin elimination the restrictions for $x_{ij}$ can be
expressed as interval constraints:
%
%
\begin{equation}
l_{ij} \le x_{ij} \le u_{ij},\label{eq4}
\end{equation}
where $l_{ij}$ may be $- \infty$ and $u_{ij}$ may be $\infty$.

We have such an interval constraint (\ref{eq4}) for each record $i$ in which the
value of variable $x_{j}$ is missing. Now, the problem for variable $x_{j}$
is to fill in the missing values with imputations, such that the sum
constraint for variable $x_{j}$ and the interval constraints (\ref{eq4}) are
satisfied. For this we will use one of our sequential imputation algorithms
(see below). As an alternative to using these sequential imputation
algorithms for benchmarking to sum constraints, one could consider using (a
generalized version of) the approach of
\citet{KimHon12}.

When used for automatic detection of errors, Fourier--Motzkin elimination
and the related Fellegi--Holt approach can be very time-consuming to apply.
As argued in Coutinho, De Waal and Remmerswaal (\citeyear{CoudeWRem11}) and Coutinho, De Waal
and Shlomo (\citeyear{CoudeWShl13}), this is much less so for the case of
imputation.

\begin{example}\label{ex1}
To illustrate how a sequential approach can be
used, we consider a case where we have $r$ records with only three
variables as shown in Table~\ref{tab1}.

\begin{table}
\caption{Illustration of a data set}\label{tab1}
\begin{tabular*}{250pt}{@{\extracolsep{\fill}}lll@{}}
\hline
$x_{11}$ & $x_{12}$ & $x_{13}$\\
$x_{21}$ & $x_{22}$ & $x_{23}$\\
$\vdots$ & $\vdots$ & $\vdots$\\
$x_{r1}$ & $x_{r2}$ & $x_{r3}$\\[6pt]
$X_{1}$ & $X_{2}$ & $X_{3}$\\
\hline
\end{tabular*}
\end{table}

These columns contain missing values that require imputation. Suppose that
the data have to satisfy the following edit restrictions:
%
%
\begin{eqnarray}
x_{i1} + x_{i2} &=& x_{i3}, \label{eq5}
\\
x_{i1} &\geq& x_{i2}, \label{eq6}
\\
x_{i3}& \geq& 3x_{i2}, \label{eq7}
\\
x_{ij} &\geq& 0\qquad (j=1,2,3). \label{eq8}
\end{eqnarray}
In addition, suppose that the following population total restrictions have
to be satisfied:
%
%
\begin{equation}
\sum_{i = 1}^{r} x_{ij} =
X_{j}\qquad (j=1,2,3). \label{eq9}
\end{equation}

We select a variable to be imputed, say, $x_{3}$. Suppose that the observed
value of variable $x_{1}$ in record $i$ equals 10 and the values of
variables $x_{2}$ and $x_{3}$ are missing for that record. The reduced set
of edits for record $i$ is then given by
%
%
\begin{eqnarray}
10 + x_{i2} &=& x_{i3},\label{eq10}
\\
10 &\geq& x_{i2},\label{eq11}
\\
x_{i3} &\geq& 3x_{i2},\label{eq12}
\\
x_{ij} &\geq& 0 \qquad(j=2,3). \label{eq13}
\end{eqnarray}

We eliminate $x_{i2}$ by substituting the expression $x_{i2} = x_{i3}-10$ into the other edits (\ref{eq11}) to (\ref{eq13}). We obtain the following set of
inequalities for $x$\tsub{$i3$}:
%
%
\begin{eqnarray}
x_{i3} &\geq& 3(x_{i3} - 10), \label{eq14}
\\
x_{i3} - 10 &\geq& 0. \label{eq15}
\end{eqnarray}
Once we have obtained an imputation value for $x$\tsub{$i3$,} we can obtain
a value for $x_{i2}$ satisfying all edits by filling in the imputation
value for $x_{i3}$ into (\ref{eq10}).

In this case there are no remaining variables except $x_{i3}$ itself, so
Fourier--Motzkin elimination is not needed anymore. Inequality (\ref{eq14}) is
obviously equivalent to $x_{i3} \leq$ 15 and (\ref{eq15}) to $x_{i3} \geq$ 10, so
the admissible interval for $x_{3}$ for record~$i$ is given by $10 \leq
x_{i3} \leq$ 15. After we have obtained interval constraints for $x_{3}$
for each record in which the value of $x_{3}$ is missing, we impute values
for $x_{3}$ in all these records by means of one of our sequential
imputation algorithms (see below).
\end{example}

\subsection{Adjusted predicted mean imputation}\label{sec3.2}

In the previous subsection we explained how a sequential approach can be
used. Now we are ready to describe our imputation algorithms. The idea of
the first algorithm is to obtain predicted mean imputations that satisfy
the sum constraint and then adjust these imputations such that they also
satisfy the interval constraints. To illustrate this idea, we use a simple
regression model with one predictor but generalization to regression models
with multiple predictors is straightforward.

\subsubsection{Standard regression imputation}\label{sec3.2.1}

Suppose that we want to impute a target column $\mathbf{x}_{t}$, that is,
the column vector with (possibly missing) values $x_{it}\ (i=1,\ldots,r)$
using as a predictor a column $\mathbf{x}_{p}$. The standard regression
imputation approach is based on the model
%
%
\begin{equation}
\mathbf{x}_{t} = \beta_{0}\mathbf{1} + \beta
\mathbf{x}_{p} + \bolds{\varepsilon},\label{eq16}
\end{equation}
where \textbf{1} is the vector of appropriate length with ones in every
entry and $\bolds{\varepsilon}$ is a vector with random residuals.

We assume that the predictor is either completely observed or already
imputed, so there are no missing values in the predictor (anymore). There
are of course missing values in $\mathbf{x}_{t}$ and to estimate the model,
we can only use the records for which both $\mathbf{x}_{t}$ and
$\mathbf{x}_{p}$ are observed. The data matrix for estimation consists of
the columns $\mathbf{x}_{t.\mathrm{obs}},\mathbf{x}_{p.\mathrm{obs}}$, where {obs}
denotes the records with $\mathbf{x}_{t}$ observed (and {mis} will
denote the opposite). Under the assumption of MAR, we can use ordinary
least squares (OLS) estimators of the parameters, $\hat{\beta}_{0}$ and
$\hat{\beta}$, to obtain predictions for the missing values in
$\mathbf{x}_{t}$:
\[
\hat{\mathbf{x}}_{t.\mathrm{mis}} = \hat{\beta}_{0}\mathbf{1} + \hat{
\beta} \mathbf{x}_{p.\mathrm{mis}},
\]
where $\mathbf{x}_{p.\mathrm{mis}}$ contains the $\mathbf{x}_{p}$-values for the
records with $\mathbf{x}_{t}$ missing and $\hat{\mathbf{x}}_{t.\mathrm{mis}}$ are
the predictions for the missing $\mathbf{x}_{t}$-values in those records.
The imputed column $\tilde{\mathbf{x}}_{t}$ consists of the observed values
and the predicted values filled in for the missing values
$\tilde{\mathbf{x}}_{t} =
(\mathbf{x}_{t.\mathrm{obs}}^{T},\hat{\mathbf{x}}_{t.\mathrm{mis}}^{T})^{T}$, where the
superscript $T$ denotes the transpose.

These imputed values will generally not satisfy the sum constraint, but a
slightly modified regression approach can ensure that they do and will be
described next.

\subsubsection{Extending the standard regression imputation to satisfy the
sum-constraint}\label{sec3.2.2}

To describe the extended regression model, we consider the following model
for the target variable that differentiates between observed and missing
values:
%
%
\begin{equation}
\pmatrix{\mathbf{x}_{t.\mathrm{obs}}
\cr
\mathbf{x}_{t.\mathrm{mis}}} = \pmatrix{ 1
& 0 & \mathbf{x}_{p.\mathrm{obs}}
\cr
0 & 1 & \mathbf{x}_{p.\mathrm{mis}}}\pmatrix{
\beta_{0}
\cr
\beta_{1}
\cr
\beta} + \pmatrix{ \bolds{
\varepsilon}_{\mathrm{obs}}
\cr
\bolds{\varepsilon}_{\mathrm{mis}} }.
\label{eq17}
\end{equation}

Apart from a coefficient $\beta$ for the predictor $\mathbf{x}_{p}$, the
model consists of two separate constants (coefficients $\beta_{0}$ and
$\beta_{1}$), one for the observed values in the target variable and one
for the missing ones. This model cannot be estimated because
$\mathbf{x}_{t.\mathrm{mis}}$ is missing. However, the total of these missing values
is known because we have assumed that the total of the target variable
$X_{t}$ is known and, hence, the total of the missing values is $X_{t.\mathrm{mis}}
= X_{t} - \sum_{i} x_{t.\mathrm{obs}.i}$. For the data that we actually observe, the
model is (by summing over the, say, $m$ units with missing values in the
target variable)
%
%
\begin{eqnarray}\label{eq18}
\pmatrix{ \mathbf{x}_{t.\mathrm{obs}}
\cr
X_{t.\mathrm{mis}} } &=& \pmatrix{ 1 & 0 &
\mathbf{x}_{p.\mathrm{obs}}
\cr
0 & m & X_{p.\mathrm{mis}} }\pmatrix{
\beta_{0}
\cr
\beta_{1}
\cr
\beta } + \pmatrix{ \bolds{
\varepsilon}_{\mathrm{obs}}
\cr
0 } \quad\mbox{or}
\nonumber
\\[-8pt]
\\[-8pt]
\nonumber
\mathbf{y} &=& \mathbf{Z}\bolds{
\beta} + \bolds{\varepsilon}\qquad \mathrm{say},
\end{eqnarray}
with $X_{p.\mathrm{mis}} = \sum_{i} x_{p.\mathrm{mis}.i}$. Notice the zero residual in the
equation corresponding to $X_{t.\mathrm{mis}}$, reflecting the requirement that the
predicted value of the sum $X_{t.\mathrm{mis}}$ should equal the known observed
value.

If the parameter vector $\bolds{\beta}$ in (\ref{eq18}) is estimated by
applying OLS to the data $\mathbf{y}$ and~$\mathbf{Z}$, the estimator
$\hat{\bolds{\beta}}$ will solve the normal equations
$\mathbf{Z}^{T}(\mathbf{y} - \hat{\mathbf{y}}) = \mathbf{0}$, with
components corresponding to the columns of $\mathbf{Z}$,
\begin{subequations}
\begin{eqnarray}
\mathbf{1}^{T}(\mathbf{x}_{t.\mathrm{obs}} - \hat{\mathbf{x}}_{t.\mathrm{obs}})
&=& 0,\label{eq19a}
\\
m(X_{t.\mathrm{mis}} - \hat{X}_{t.\mathrm{mis}})& =& 0,\label{eq19b}
\\
\mathbf{x}_{p.\mathrm{obs}}^{T}(\mathbf{x}_{t.\mathrm{obs}} - \hat{
\mathbf{x}}_{t.\mathrm{obs}}) + X_{p.\mathrm{mis}}(X_{t.\mathrm{mis}} -
\hat{X}_{t.\mathrm{mis}}) &=& 0. \label{eq19c}
\end{eqnarray}
\end{subequations}

From (\ref{eq19b}) we obtain $\hat{X}_{t.\mathrm{mis}} = X_{t.\mathrm{mis}}$, which shows that the
sum of the estimated predictions indeed equals its known value.
Furthermore, by substituting this result in (\ref{eq19c}), we obtain
$\mathbf{x}_{p.\mathrm{obs}}^{T}(\mathbf{x}_{t.\mathrm{obs}} - \hat{\mathbf{x}}_{t.\mathrm{obs}}) =
0$. Thus, (\ref{eq19a}) and (\ref{eq19c}) do not involve the totals $X_{t.\mathrm{mis}}$ and
$X_{p.\mathrm{mis}}$ and are equal to the normal equations for the standard
regression model (\ref{eq16}), fitted on the data
$\mathbf{x}_{t.\mathrm{obs}},\mathbf{x}_{p.\mathrm{obs}}$. Consequently, the parameter
estimates corresponding to (\ref{eq19a}) and (\ref{eq19c}), $\hat{\beta}_{0}$ and
$\hat{\beta}$, are equal to the parameter estimates obtained for the
standard model. The parameter estimate $\hat{\beta}_{1}$ adds to this model
a constant for the predicted missing values\vadjust{\goodbreak} such that the sum constraint is
satisfied. Using the estimates $\hat{\beta}_{1}$ and $\hat{\beta}$, the
missing values are imputed by the predicted values according to (\ref{eq17}):
\[
\hat{\mathbf{x}}_{t.\mathrm{mis}} = \hat{\beta}_{1}\mathbf{1} + \hat{
\beta} \mathbf{x}_{p.\mathrm{mis}}.
\]

In the case of unequal weights, the regression method described above must
be modified to take these weights into account. First, to obtain a (design)
consistent estimator of $\bolds{\beta}$, weighted least squares should
be applied with weights equal to the design or calibration weights $w_{i}$
(see Section~\ref{sec2.1}). Second, the summation over the units with missing values
that lead to (\ref{eq18}) will now be replaced by a weighted summation which leads
to redefining the following quantities: $X_{t.\mathrm{mis}} = \sum_{i}
w_{i}x_{t.\mathrm{mis}.i} = X_{t}^{\mathrm{pop}} - \sum_{i} w_{i}x_{t.\mathrm{obs}.i}$, with $m$ the
sum of the weights of the missing units rather than the number of missing
units and $X_{p.\mathrm{mis}} = \sum_{i} w_{i}x_{p.\mathrm{mis}.i}$. With these
modifications, parameters $\hat{\beta}_{1}$ and $\hat{\beta}$ obtained from
WLS estimation of the model (\ref{eq18}) can be used for imputation as before but
now resulting in imputations that satisfy the \textit{weighted} sum
constraint.

\subsubsection{Adjusting regression imputations to satisfy interval
constraints}\label{sec3.2.3}
$\!\!\!$ Since the interval constraints have not been considered in obtaining the
predicted values, it can be expected that a number of these predictions are
not within their admissible intervals. One way to remedy this situation is
to calculate adjusted predicted values defined by
\[
\hat{\mathbf{x}}_{t.\mathrm{mis}}^{\mathrm{adj}} = \hat{\mathbf{x}}_{t.\mathrm{mis}}
+ \mathbf{a}_{t},
\]
with $\mathbf{a}_{t}$ a vector with adjustments to be added to the
predictions such that the adjusted predictions satisfy both the sum
constraint (which is equivalent to \mbox{$\sum_{i} a_{t,i} = 0$}) and the interval
constraints, and the adjustments are as small as possible. One way to find
such a vector $\mathbf{a}_{t}$ is to solve the quadratic programming
problem
\[
\mbox{minimize } \mathbf{a}_{t}^{T}\mathbf{a}_{t}\mbox{ subject to}    \qquad
\mathbf{1}^{T}\mathbf{a}_{t} = 0\mbox{ and }\mathbf{l}_{t} \le
\hat{\mathbf{x}}_{t.\mathrm{mis}} + \mathbf{a}_{t} \le \mathbf{u}_{t},
\]
with $\mathbf{l}_{t}$ a vector with the lower bounds and $\mathbf{u}_{t}$ a
vector containing the upper bounds. For cases with unequal weights, as
discussed in Sections~\ref{sec2.1} and \ref{sec3.2.2}, the weighted sum of the adjustments
should be zero, leading to the constraint $\mathbf{w}^{T}\mathbf{a}_{t} =
0$ instead of $\mathbf{1}^{T}\mathbf{a}_{t} = 0$, with $\mathbf{w}$ the
vector with weights for the units with missing values.

A simple algorithm for solving convex optimization problems with interval
constraints is described by \citet{CenLen81}. In our case their
iterative approach results in an algorithm that is very easy to implement.
To describe this algorithm, we first decompose the adjustments $a_{t,i}$ as
$a_{t,i} = b_{t,i} - \bar{b}_{t}$. The $b_{t,i}$ will be determined such
that the interval constraints are satisfied and $\bar{b}_{t}$ is the mean
of the $b_{t,i}$. Subtracting $\bar{b}_{t}$ from the $b_{t,i}$ ensures that
the $a_{t,i}$ sum to zero. The algorithm now proceeds as follows.
Initialize $b_{t,i} = 0$ and $\bar{b}_{t}$ and then calculate new values
for $b_{t,i}$ and update $\bar{b}_{t}$ according to the following iterative
scheme (with $j$ the iteration counter):

\begin{longlist}[1.]
\item[1.] For each $i$, find the smallest (in absolute value) possible
value for $b_{t,i}^{(j)}$ such that the interval constraint $l_{t,i} \le
\hat{x}_{t.\mathrm{mis},i} + b_{t,i}^{(j)} - \bar{b}_{t}^{(j - 1)} \le u_{t,i}$ is
satisfied.
\item[2.] Set $\bar{b}_{t}^{(j)}$ equal to the mean of $b_{t,i}^{(j)} -
\bar{b}_{t}^{(j - 1)}$.
\end{longlist}

When these two steps are iterated until convergence, that is, until the
change in the $b_{t,i}^{(j)}$ becomes negligible, the resulting
$a_{t,i}^{(j)} = b_{t,i}^{(j)} - \bar{b}_{t}^{(j - 1)}$ solve the quadratic
programming problem defined above.

We will refer to this method as \textit{BPMA} (Benchmarked Predictive Mean
imputation with Adjustments to imputations so they satisfy interval
constraints). We will also evaluate this method without benchmarking to
totals. We will refer to that method as \textit{UPMA} (Unbenchmarked
Predictive Mean imputation with Adjustments to imputations so they satisfy
interval constraints).

\subsection{Regression imputation with random residuals}\label{sec3.3}

It is well known that, in general, predictive mean imputations show less
variability than the true values that they are replacing. In order to
better preserve the variance of the true data, random residuals can be
added to the predicted means. The adjusted predictive mean imputations
considered in the previous section will also be hampered by this drawback
because these adjustments are intended to be as close as possible to the
predicted means and not to reflect the variance of the original data.

In order to better preserve the variance of the true data, we start with
the predicted values $\hat{\mathbf{x}}_{t.\mathrm{mis}}$ obtained from (\ref{eq17}) that
already satisfy the sum constraint, and our purpose is to add random
residuals to these predicted means such that the distribution of the data
is better preserved and, in addition, both the interval and sum constraints
are satisfied. These residuals serve the same purpose (satisfying the
constraints) as the adjustments $a_{ t, i}$ but, in contrast to the $a_{ t,
i}$, they are not as close as possible to the predicted means and are
intended to reflect the true variability around these predicted means.

A simple way to obtain residuals is to draw each of the $m$ residuals by
Acceptance/Rejection (AR) sampling [see, e.g., \citet{RobCas99}
for more on AR sampling] from a normal distribution with mean zero and
variance equal to the residual variance of the regression model, that is,
by repeatedly drawing from this normal distribution until a residual is
drawn that satisfies the interval constraint.

The residuals obtained by this AR sampling may not sum to zero so that the
imputed values do not satisfy the sum constraint. We may then adjust these
residuals, as little as possible, such that they sum to zero and the
interval constraints remain satisfied by applying the iterative method
described in Section~\ref{sec3.2.3}. We will refer to this method as \textit{BPMR}
(Benchmarked Predictive Mean imputation with random Residuals).

Note that in all imputation methods described in Section~\ref{sec3} (\textit{BPMA},
\textit{UPMA} and \textit{BPMR}) one can use the imputed values as
predictors. In our simulation study and evaluation study described in
Sections~\ref{sec5} and \ref{sec6} we have passed through the variables in need of imputation
multiple times in order to preserve correlations as well as possible.

\section{MCMC approach}\label{sec4}

The final imputation algorithm we describe is based on a Monte Carlo Markov
Chain [MCMC; see, e.g., \citet{RobCas99} and \citet{Liu01} for more
on MCMC in general] approach to which we will refer as \textit{MCMC}. This
MCMC approach is an extended version of similar approaches by \citet{Ragetal01}, \citet{Rub03},
Tempelman [(\citeyear{Tem07}); Chapter~6] and \citet{VanGro11}. \citet{Ragetal01} and \citet{Rub03} do
not take edits or totals into account in their MCMC approaches, while Van
Buuren and Groothuis-Oudshoorn (\citeyear{VanGro11}) take only some simple edits, such as
univariate range checks, into account and again no benchmarking to totals.
The MCMC approach of \citet{Tem07} does take edits into account, but not
totals.

Our approach starts with a fully imputed data set consistent with the edits
and known totals, for instance, obtained by means of the imputation methods
\textit{BPMA} or \textit{BPMR} described in Sections~\ref{sec3.2} and \ref{sec3.3}.
Subsequently, we try to improve the imputed values so they preserve the
statistical distribution of the data better. Our algorithm, which is
similar to so-called data swapping for categorical data [see Dalenius and
Reis (\citeyear{DalRei82})], is sketched below.

As mentioned, we start with a pre-imputed data set $D$ consistent with the
edits and known totals. We randomly select two records $s$ and $t$ from $D$
with at least one common variable $x_{j}$ with missing values in both
records. The imputed values in records $s$ and $t$ are treated as unknowns.
Next, we construct the set of edits and sum constraints that have to hold
for these unknowns. We obtain the edits for the unknowns in records $s$ and
$t$ by filling in the observed values in these records into the edits. The
sum constraints are obtained by noting that the imputed values of a certain
variable in records $s$ and $t$ should equal the known total for this
variable minus the observed values (if any) in records $s$ and $t$ and the
values of the (observed and imputed) values in the other records. We will
re-impute $x_{sj}$ and later derive the value of $x_{tj}$ in the record $t$
by subtracting the value of $x_{sj}$ from the known sum for these two
values. In this process, the values of the other imputed variables in
records $s$ and $t$ may, in principle, be changed too.

We determine an admissible interval for $x_{\mathit{sj}}$ by eliminating
all unknowns except $x_{sj}$ from the set of edits and sum constraints for
the unknowns in records $s$ and $t$ by means of Fourier--Motzkin
elimination. We draw a value for $x_{sj}$ from a posterior predictive
distribution implied by a linear regression model under an uninformative
prior, conditional on the fact that this value has to lie inside its
admissible interval. In our implementation of the algorithm we calculate
new values for the regression parameters for each pair of records. For
different variables $x_{j}$, different linear regression models, and hence
different posterior predictive distributions, are used.

If necessary for satisfying the edits and sum constraints for the unknowns
in records $s$ and $t$, we apply back-substitution, using the new imputed
value for $x_{sj}$, the sum constraints and the equations among the edits
for the unknowns, to adjust the values of the other imputed values in
records $s$ and $t$. If imputed values in records $s$ and $t$ do not have
to be adjusted, we retain the current values. Finally, we update data set
$D$ with the modified imputed values. If the distribution of the imputed
values has converged, we terminate the algorithm. Otherwise, we again
select two records with a common variable with missing values in both
records and repeat the procedure.

Note that ``convergence'' is a difficult concept, as we are referring to
the convergence of a statistical distribution. We refer to \citet{RobCas99} and
 \citet{Liu01} for more on convergence of MCMC processes.
Also note that the algorithm may not converge. In fact, convergence may not
even be possible, as the existence of a multivariate distribution that is
compatible with the various univariate posterior predictive distributions
is not guaranteed. This is a well-known theoretical drawback of such an
MCMC approach. \citet{Rub03} refers to this phenomenon as ``incompatible
MCMC.'' In practice, one usually observes the distribution of the imputed
data set over a large number of iterations and monitors whether the
observed distribution appears to converge. We have applied this pragmatic
approach as well.

An important reason why we use a posterior predictive distribution implied
by a linear regression model under an uninformative prior is that this, in
principle, allows us to extend our approach to multiple imputation. The
extension to multiple imputation is not studied in the present article,
however.

\begin{example}\label{ex2}
We illustrate some aspects of our
\textit{MCMC} approach by means of a simple example. Let us assume that
there are five variables $x_{j}\ (j=1,\ldots,5)$ and six edits given by
%
\begin{eqnarray}
x_{1} + x_{2} + x_{3} + x_{4} &=&
x_{5}, \label{eq20}
\\
x_{j}& \ge& 0 \qquad(j=1,\ldots,5). \label{eq21}
\end{eqnarray}
Let us also assume that the values of $x_{1}$ have to sum up to 10,000, of
$x_{2}$ to 12,000, of $x_{3}$ to 8000, of $x_{4}$ to 32,000, and of
$x_{5}$ to 62,000.

We start with a fully imputed data set $D$. We select two records with at
least one common variable---variable $x_{5}$ in our example---with
missing values in two records: a record $s$ where the values of variables
$x_{1}$ and $x_{2}$ were observed, say, $x_{s1} = 10$ and $x_{s2} = 15$,
and the values of variables $x_{3}$, $x_{4}$ and $x_{5}$ were missing, and
a record $t$ where the values of variables $x_{2}$ and $x_{3}$ were
observed, say, $x_{t2} = 30$ and $x_{t3} = 25$, and the values of variables
$x_{1}$, $x_{4}$ and $x_{5}$ were missing. Let us assume that the imputed
values for records $s$ and $t$ in data set $D$ are given by $x_{s3} = 20$,
$x_{s4} = 30$, $x_{s5} = 75$, $x_{t1} = 15$, $x_{t4} = 35$ and $x_{t5} =
105$. As $D$ is consistent with the edits and known totals, the sum of the
imputed and observed values in the other records hence must equal 9975 for
variable $x_{1}$, 11,955 for variable $x_{2}$, 7955 for variable $x_{3}$,
31,935 for variable $x_{4}$ and 61,820 for variable $x_{5}$. The situation
for records $s$ and $t$ is summarized in Table~\ref{tab2}, where a ``?'' means that
the corresponding value has been imputed and may be re-imputed.

\begin{table}
\caption{Situation for records $s$ and $t$}
\label{tab2}
\begin{tabular*}{\textwidth}{@{\extracolsep{\fill}}lccccc@{}}
\hline
& $\bolds{x_{1}}$ & $\bolds{x_{2}}$ & $\bolds{x_{3}}$ & $\bolds{x_{4}}$ & $\bolds{x_{5}}$\\
\hline
$s$ & 10 & 15 & ? & ? & ?\\
$t$ & ? & 30 & 25 & ? & ?\\[6pt]
Total for $s$ and $t$ & 25 & 45 & 45 & 65 & 180\\
\hline
\end{tabular*}
\end{table}

We fill in the observed values in both records into the edits and obtain
$25 + x_{s3} + x_{s4} = x_{s5}$, and $x_{sj} \ge 0\ (j=3,4,5$) for record
$s$, and $55 + x_{t1} + x_{t4} = x_{t5}$ and $x_{tj} \ge 0\ (j=1,4,5$) for
record $t$. The sum constraints for the unknowns in records $s$ and $t$ are
given by $x_{t1} = 15$, $x_{s3} = 20$, $x_{s4} + x_{t4} = 65$, $x_{s5} +
x_{t5} = 180$.

We eliminate all unknowns except $x_{s5}$ from the above set of
constraints. We obtain the interval $45 \le x_{s5} \le 110$. We draw a
value for $x_{s5}$ from a posterior predictive distribution implied by a
linear regression model under an uninformative prior, conditional on the
fact that $45 \le x_{s5} \le 110$, say, we draw the value 100 for $x_{s5}$.
Finally, we use back-substitution to obtain adjusted imputed values:
$x_{s4} = 55$, $x_{t4} = 10$ and $x_{t5} = 80$.

We update data set $D$ with the adjusted imputed values and check whether
the distribution has converged. If so, we terminate the algorithm.
Otherwise, we repeat the procedure.
\end{example}

\section{Simulation study}\label{sec5}

A simulation study was carried out emulating the 2005 Israel Income Survey
used in the evaluation study as presented in Section~\ref{sec6}. For this design,
although stratified sampling was employed, every individual had the same
inclusion probability. Therefore, the results from the simulation study can
be viewed as arising from a single stratum. We generated variables $x_{1}$,
$x_{2}$ and a predictor $P$ from a normal distribution using linear
transformations to ensure a reasonably realistic degree of correlation
between them. The simulated population data set included 100,000 records.
The means for $x_{1}$ and $x_{2}$ in the population are 3902 and 991 and
standard deviations 636 and 401, respectively. The correlation between
$x_{1}$ and $x_{2}$ is 0.87, between $x_{1}$ and $P$ 0.66 and between
$x_{2}$ and $P$ 0.57. Edit constraints (\ref{eq5}) to (\ref{eq8}) and sum constraint (\ref{eq9})
are all preserved on the simulated population data set, where $P$ is
variable $x_{3}$ in (\ref{eq5}) to (\ref{eq9}). Out of the 100,000 records in the
population data set, 20,000 (20\%) records were randomly selected and their
$x_{1}$ variable blanked out. Half of those selected records also had their
$x_{2}$ variable blanked out. An additional 10\% of the remaining records
were randomly selected and their $x_{2}$ variable blanked out. This
represents a MCAR nonresponse mechanism.

The simulation study is based on drawing 1:20 random samples from the
population, that is, the sample size is $n=5000$, and the imputation
procedures applied are as outlined in Sections~\ref{sec3} and \ref{sec4}:

\begin{itemize}
\item \textit{UPMA}---unbenchmarked simple predictive mean
imputation (Sec-\break tion~\ref{sec3.2.1}) with adjustments to imputations so they satisfy
interval constraints (Section~\ref{sec3.2.3}). In this method the only stochastic
effects are from the estimation of the parameters in model (\ref{eq17}).
\item \textit{BPMA}---benchmarked predictive mean imputation
(Section~\ref{sec3.2.2}) with adjustments to imputations so they satisfy interval
constraints (Sec-\break tion~\ref{sec3.2.3}). Again, in this method the only stochastic
effects are from the estimation of the parameters in model (\ref{eq17}).
\item \textit{BPMR}---benchmarked predictive mean imputation
(Section~\ref{sec3.2.2}) with random residuals (Section~\ref{sec3.3}). In this method there
is an extra stochastic effect in comparison to \textit{UPMA} and
\textit{BPMA} due to the addition of random residuals.
\item \textit{MCMC}---the approach described in Section~\ref{sec4}. The
data set with \textit{BPMA} was used as the pre-imputed data set for our
\textit{MCMC} approach. In this method there are extra stochastic effects
in comparison to \textit{UPMA} and \textit{BPMA} due to selecting pairs of
records and drawing new values for some of the fields in those records.
\end{itemize}

\begin{table}[b]
\tabcolsep=0pt
\caption{Average mean and standard deviation of x $_{1}$
and x $_{2}$ from 300 samples}
\label{tab3}
\begin{tabular*}{\textwidth}{@{\extracolsep{\fill}}lccccccc@{}}
\hline
& \multicolumn{2}{c}{$\bolds{x_{1}}$} &
\multicolumn{2}{c}{$\bolds{x_{2}}$} & \multicolumn{3}{c@{}}{\textbf{Correlations}}\\[-6pt]
& \multicolumn{2}{c}{\hrulefill} &
\multicolumn{2}{c}{\hrulefill} & \multicolumn{3}{c@{}}{\hrulefill}\\
{\textbf{Method}} & \textbf{Mean} & \textbf{Standard deviation} & \textbf{Mean} & \textbf{Standard deviation} & $\bolds{x_{1}}$\textbf{,} $\bolds{x_{2}}$
& $\bolds{x_{1}}$\textbf{,} $\bolds{P}$ & $\bolds{x_{2}}$\textbf{,} $\bolds{P}$\\
\hline
Original & 3902 & 635 & 991 & 400 & 0.87 & 0.66 & 0.57\\
\textit{UPMA} & 3901 & 599 & 991 & 382 & 0.86 & 0.70 & 0.60\\
\textit{BPMA} & 3902 & 599 & 991 & 382 & 0.86 & 0.70 & 0.60\\
\textit{BPMR} & 3902 & 637 & 991 & 393 & 0.79 & 0.66 & 0.58\\
\textit{MCMC} & 3902 & 692 & 991 & 416 & 0.76 & 0.60 & 0.54\\
\hline
\end{tabular*}
\end{table}

We repeated the sampling 300 times in order to investigate the impact of
the imputation procedures on the sample distribution. We also computed the
average across the samples of some commonly used evaluation metrics for
comparing imputation procedures [\citet{Cha03}, \citet{PanDeW05}]. These include the following:

\begin{itemize}
\item $d_{L1}$ measure: $d_{L1} = \frac{\sum_{i \in M}
w_{i}\llvert  \hat{x}_{i} - x_{i}^{*} \rrvert } {\sum_{i \in M} w_{i}}$, where
$\hat{x}_{i}$ is the imputed value in record $i$ and $x_{i}^{*}$ is the
original value of the variable, $M$ denotes the set of $m$ records with
imputed values for variable $x$ and $w_{i}$ is the raising weight for
record~$i$.
\item \textit{K--S} Kolmogorov--Smirnov test statistic to compare the
empirical distribution of the original values to the empirical distribution
of the imputed values, $\mbox{\textit{K--S}} = \max_{j}(|F_{x^{*}}(t_{j}) -
F_{\hat{x}}(t_{j})|)$, where the $\{ t_{j}\}$ values are the $2m$ jointly
ordered original and imputed values of $x$, and $F_{x^{*}}$ and $F_{\hat{x}}$ denote the empirical distributions of the original and imputed values, respectively.
\item The percent difference between the standard deviation
(STD) of $x_{1}$ and $x_{2}$ in the sample data with imputations to the
standard deviation of the original sample data:
\end{itemize}
\[
100\frac{(\mathit{STD}_{\mathrm{imp}} - \mathit{STD}_{\mathrm{orig}})}{\mathit{STD}_{\mathrm{orig}}}.
\]

For all methods, the variable $x_{1}$ was first regressed on the predictor
$P$, and $x_{2}$ was first regressed on the predictor $P$ and $x_{1}$. In
our study, we use the imputation methods \textit{UPMA}, \textit{BPMA} and
\textit{BPMR} in an iterative way, as mentioned at the end of Section~\ref{sec3}.
That is, after all variables have been imputed once, the following rounds
of the procedure uses, for each variable to be re-imputed, all other
variables as predictors. Thus, after the first round $x_{1}$ is regressed
on $P$ and $x_{2}$, and $x_{2}$ is regressed on $P$ and $x_{1}$. The
regression model for the \textit{MCMC} method is based on the sequential
regression model of \citet{Ragetal01} and drawing values from the
corresponding predictive distributions. Table~\ref{tab3} examines the impact of the
imputation on the sample distribution by comparing the original mean,
standard deviation and correlations in the population data set with the
average mean, Monte Carlo standard deviation and correlations obtained from
the 300 samples. Table~\ref{tab4} contains the average of the evaluation metrics
used to assess the imputation methods across the 300 samples. Note that
\textit{UPMA} and \textit{BPMA} are deterministic imputations and
\textit{BPMR} and \textit{MCMC} stochastic ones.

\begin{table}
\tabcolsep=0pt
\caption{Average evaluation metrics for the imputation methods from
300 samples}
\label{tab4}
\begin{tabular*}{\textwidth}{@{\extracolsep{\fill}}ld{3.3}d{3.3}d{3.3}d{3.3}d{3.3}d{3.3}d{3.3}d{3.3}@{}}
\hline
& \multicolumn{4}{c}{$\bolds{x_{1}}$} &
\multicolumn{4}{c@{}}{$\bolds{x_{2}}$}\\[-6pt]
& \multicolumn{4}{c}{\hrulefill} &
\multicolumn{4}{c@{}}{\hrulefill}\\
& \multicolumn{1}{c}{\textbf{\textit{UPMA}}} & \multicolumn{1}{c}{\textbf{\textit{BPMA}}} & \multicolumn{1}{c}{\textbf{\textit{BPMR}}} &
\multicolumn{1}{c}{\textbf{\textit{MCMC}}} &
\multicolumn{1}{c}{\textbf{\textit{UPMA}}} & \multicolumn{1}{c}{\textbf{\textit{BPMA}}} & \multicolumn{1}{c}{\textbf{\textit{BPMR}}} &
\multicolumn{1}{c@{}}{\textbf{\textit{MCMC}}}\\
\hline
Distance $d_{L1}$ & 382 & 382 & 535 & 640 & 206 & 206 & 270 & 380\\
Kolmogorov--&&&&&&&&\\
\quad Smirnov \textit{K--S} & 0.116 & 0.113 & 0.030 & 0.075 & 0.145 &
0.146 & 0.098 & 0.089\\
\% difference&&&&&&&&\\
\quad of STD & -5.7\% & -5.7\% & 0.2\% & 9.0\% & -4.5\% & -4.6\% &
-1.7\% & 4.0\% \\
\hline
\end{tabular*}
\end{table}

The results in Table~\ref{tab3} show that since all the methods, except
\textit{UPMA}, benchmark to known totals, there is no bias for these
methods introduced into the imputed data. As expected with mean imputation,
the variance for the deterministic methods \textit{UPMA} and \textit{BPMA}
is reduced. While both methods preserve the edit constraints across the
individual records, the \textit{BPMA} approach benchmarks the total. Out of
the stochastic methods, \textit{BPMR} based on random residuals preserves
the variance with only a slight decrease in the correlation between $x_{1}$
and $x_{2}$. The \textit{MCMC} algorithm, however, increases the variance
and has more of a decrease in the correlation structure of the variables.

The results in Table~\ref{tab4} show the similarities between the methods
\textit{UPMA} and \textit{BPMA} with respect to the evaluation metrics.
Both methods show lower distance $d_{L1}$ and larger relative differences
to the standard deviation of the mean compared to the stochastic methods
\textit{BPMR} and \textit{MCMC} as expected with deterministic mean
imputation. In addition, the Kolmogorov--Smirnov (\textit{K--S}) statistics
are larger for the deterministic methods than the stochastic methods.
Comparing the two stochastic methods \textit{BPMR} and \textit{MCMC}, the
results in Table~\ref{tab4} show that the distance $d_{L1}$ and the relative
difference of the standard deviation of the mean are higher for the
\textit{MCMC} approach for both variables $x_{1}$ and $x_{2}$. The
\textit{MCMC} approach also has a higher \textit{K--S} statistic compared to
the \textit{BPMR} method for $x_{1}$ but slightly lower for $x_{2}$.

Our general conclusion from the simulation study is that, based on the
preservation of totals (and edit constraints), preservation of standard
deviations and preservation of other distributional properties, we consider
\textit{BPMR} to be the most promising method. This will be tested further
on a real data set in Section~\ref{sec6}.

\section{Evaluation study}\label{sec6}

\subsection{Evaluation data set}\label{sec6.1}

We use a real data set from the 2005 Israel Income Survey. The file for the
evaluation study contains 11,907 individuals aged 15 and over that
responded to all the questions in the questionnaire of the 2005 Israel
Income Survey and, in addition, earned more than 1000 Israel Shekels (IS)
for their monthly gross income. We focus on three variables from the Income
Survey: the gross income from earnings (\textit{gross}), the net income
from earnings (\textit{net}) and the difference between them
(\textit{tax}). As above, we consider the following edits for each record
$i$:
\begin{eqnarray*}
\mathit{net}_{i} + \mathit{tax}_{i} &=&
\mathit{gross}_{i},
\\
\mathit{net}_{i} &\geq& \mathit{tax}_{i},
\\
\mathit{gross}_{i} &\geq& 3 \times \mathit{tax}_{i},
\\
\mathit{gross}_{i} &\geq& 0, \qquad\mathit{net}_{i} \geq 0,\qquad \mathit{tax}_{i} \geq 0.
\end{eqnarray*}
Item nonresponse was introduced randomly to the income variables in order
to simulate a typical data set: 20\% of the records (2382 records) were
selected randomly and their net income variable blanked out. Half of those
selected records (1191 records) also had their tax variable blanked out.
An additional 10\% (1191 records) were selected randomly from the data set
and their tax variable deleted. We assume that the totals of each of the
income variables, including tax, are known.

\subsection{Evaluation results}\label{sec6.2}

The predictors that were chosen for the predictive mean imputation based on
regression modeling (\textit{UPMA}, \textit{BPMA} and \textit{BPMR}) were
the following: 14 categories of economic branch, 10 categories of
occupation, 10 categories of age group, and sex. For each category a dummy
variable was created.

In order to ensure the normality of the income variables, a log
transformation was carried out. This meant we had to change the algorithm
described in Section~\ref{sec3.2.2} slightly since the sum of the log transformed
variables which will equal the known log totals will not necessarily mean
that the sum of the original variables will equal the known original
totals. We used a correction factor to replace the constant term of the
regression to constrain the sum of the untransformed, original variables to
the original totals. We denote $\mathbf{z} = \log \mathbf{x}$, where the
logarithm is taken component-wise,\vspace*{1pt} that is, $\mathbf{z} = (\log (x_{1}),
\ldots,\log (x_{r}))$, where $r$ is the number of records. From (\ref{eq17}),
$\hat{\mathbf{z}}_{t.\mathrm{mis}} = \hat{\beta}_{1}\mathbf{1} + \hat{\beta}
\mathbf{z}_{p.\mathrm{mis}}$ and, therefore, $\hat{\mathbf{x}}_{t.\mathrm{mis}} = \exp
(\hat{\beta}_{1}) \times \exp (\hat{\beta} \mathbf{z}_{p.\mathrm{mis}})$, where
$\exp (\hat{\beta} \mathbf{z}_{p.\mathrm{mis}})$ is again taken component-wise.
Summing across the missing values gives $\hat{X}_{t.\mathrm{mis}} = \sum_{i}
\hat{x}_{t.\mathrm{mis},i} = \exp (\hat{\beta}_{1})\sum_{i} \exp ( \hat{\beta}
z_{p.\mathrm{mis},i})$. The correction replaces the constant factor $\exp
(\hat{\beta}_{1})$ with $\frac{\hat{X}_{t.\mathrm{mis}}}{\sum_{i} \exp (\hat{\beta}
z_{p.\mathrm{mis},i})}$.

Table~\ref{tab5} contains the results of the evaluation measures as described in
Section~\ref{sec5}.

\begin{table}
\tabcolsep=0pt
\caption{Results of evaluation measures for the imputation methods
in the evaluation study}
\label{tab5}
\begin{tabular*}{\textwidth}{@{\extracolsep{\fill}}ld{4.3}d{4.3}d{4.3}d{4.3}d{3.3}d{3.3}d{3.3}d{4.3}@{}}
\hline
 & \multicolumn{4}{c}{\textbf{Net income
variable}} & \multicolumn{4}{c@{}}{\textbf{Tax variable}}\\[-6pt]
 & \multicolumn{4}{c}{\hrulefill} & \multicolumn{4}{c@{}}{\hrulefill}\\
\multicolumn{1}{@{}l}{{\textbf{Evaluation measures}}}&
\multicolumn{1}{c}{\textbf{\textit{UPMA}}} & \multicolumn{1}{c}{\textbf{\textit{BPMA}}} & \multicolumn{1}{c}{\textbf{\textit{BPMR}}} &
\multicolumn{1}{c}{\textbf{\textit{MCMC}}} &
\multicolumn{1}{c}{\textbf{\textit{UPMA}}} & \multicolumn{1}{c}{\textbf{\textit{BPMA}}} &
\multicolumn{1}{c}{\textbf{\textit{BPMR}}} & \multicolumn{1}{c@{}}{\textbf{\textit{MCMC}}}\\
\hline
Distance $d_{L1}$ & 2040.4 & 2132.6 & 2695.9 & 2664.2 & 980.6 & 821.7 &
818.6 & 1154.4 \\
Kolmogorov--&&&&&&&&\\
\quad Smirnov \textit{K--S} & 0.098 & 0.149 & 0.049 & 0.086 & 0.433 &
0.323 & 0.184 & 0.155 \\
\% difference&&&&&&&&\\
\quad to STD & -41.1\% & -37.6\% & -11.9\% & -19.4\% & -3.2\% &
-4.7\% & -3.2\% & 3.5\% \\
\hline
\end{tabular*}
\end{table}

From the results of Table~\ref{tab5}, the \textit{BPMA} approach and the stochastic
approaches \textit{BPMR} and \textit{MCMC} all preserve the totals in the
data, as they should. The results on the $d_{L1}$ measure are mixed, with
the net income variable doing slightly worse for both stochastic approaches
but the tax variable showing improvement compared to the \textit{BPMR}
approach. The distribution is preserved better for the stochastic
approaches as reflected in the \textit{K--S} statistic and the percent
difference in the standard deviation of the mean. The measures when
benchmarking the totals (\textit{BPMA}) appear to be mixed compared to not
benchmarking (\textit{UPMA}) depending on the variable.

It is more difficult to draw general conclusions for the real data set than
it was for the simulated data set, since the results for the real data set
are not univocal across variables. However, based on the fact that the
stochastic methods preserve totals (and edit constraints) and preserve
standard deviations and other distributional properties better than
\textit{UPMA} and \textit{BPMA}, we consider \textit{BPMR} and
\textit{MCMC} the most promising methods.

\section{Discussion}\label{sec7}

In this article we have proposed three imputation methods for numerical
data that satisfy edit restrictions and preserve totals. Two of the
developed methods are stochastic, aiming to better preserve the variation
in the imputed data.

In this article we have not examined variance estimation after imputation.
In general, there are three approaches to variance estimation with imputed
data [see Haziza (\citeyear{Ha09}), and Chapter~7
in De Waal, Pannekoek and Scholtus (\citeyear{DeWPanSch11})]:

\begin{itemize}
\item \textit{The analytical approach}. In the analytic approach
explicit formulas are derived for variance estimation after imputation.
These formulas can be seen as adding a correction term to standard variance
formulas to take the fact that imputation is used into account. Such
formulas have been derived for standard regression imputation without
constraints [see, e.g., \citet{Fay91}, S\"{a}rndal (\citeyear{Sar92}), Deville and
S\"{a}rndal (\citeyear{DevSar94}), \citet{RaoSit95}, \citet{ShaSte99} and
\citet{Bea05}]. For our situation, where data have to satisfy edits and
population totals, analytic variance formulas have still to be developed.
\item \textit{The resampling approach}. Methods, such as the
jackknife, bootstrap and balanced repeated replication, have been used
often for variance estimation in complex surveys with imputed data [see,
e.g., \citet{Wol85}, \citet{RaoSha92},\vadjust{\goodbreak} \citet{ShaSit96} and \citet{Sha02}]. This approach is more general than the analytical approach,
because the same procedure can be used largely irrespective of the
imputation and estimation procedure that is used. Such methods could be
very well applied to the methods considered in this article.
\item \textit{Multiple imputation}. Multiple imputation was
originated by Rubin (\citeyear{Rub}, \citeyear{Rub87}). In this framework, a number of
imputations (typically 5) are obtained for each missing value and,
consequently, multiple estimates of the target parameters are obtained.
Simple formulas exist that combine the multiple estimates to a single one
and, most importantly, employ the variance between the estimates to obtain
an estimator for the variance of the combined parameter estimate. The
\textit{MCMC} method fits in the framework of multiple imputation, and
variance estimation according to Rubin's formulas for multiply imputed data
would be a natural approach for this method.
\end{itemize}

A debate about the advantages and disadvantages of the different approaches
for variance estimation after imputation and their applicability for
different purposes was published in 1996 in the Journal of the American
Statistical Association [\citet{Rub96}, \citet{Fay96}, \citet{Rao96}].

The methods introduced in this article can also be used for mass imputation
of numerical data. In \citet{Hou04} a statistical database for social
data was constructed using so-called repeated weighting based on regression
estimators. While benchmarking totals (either based on registers or
weighted survey estimates), the method does not preserve edit constraints.
The methods in this article provide an alternative to repeated weighting
which can benchmark totals, preserve edit constraints and preserve
correlation structures in the data. Initial work in the area of mass
imputation for a numerical data set having the above properties using the
methods proposed in the present article is described in
\citet{ShlDeWPan09}.\looseness=-1

From a production point of view of a statistical office, our methods are
sufficiently fast and appear to produce data of sufficiently high quality.
A~practical point of concern is the complexity of our methods. The
\textit{MCMC} method, for example, is easy to program but may be
problematic in the day-to-day routine of producing timely statistical data,
because ``convergence'' of the method is not easy to verify. For the other
methods, \textit{UPMA}, \textit{BPMA} and \textit{BPMR}, this is less of a
problem, as these methods can easily be implemented in a standard software
package.



\printaddresses

\end{document}